\documentclass[amsmath, amssymb, english, aps, prl, twocolumn, reprint,floatfix, superscriptaddress]{revtex4-1}
\usepackage[english]{babel}
\usepackage{graphicx}
\usepackage{verbatim}
\def \mb{\begin{displaymath}}
\def \me{\end{displaymath}}
\def \eb{\begin{equation}}
\def \ee{\end{equation}}
\def \Sr2RuO4{Sr${}_2$RuO$_{4}$}
\def \CaSrRuO4{Ca${}_{2-x}$Sr${}_x$RuO$_{4}$}

\def\expect#1{\mathinner{\langle{#1}\rangle}}

{\catcode`\|=\active
  \gdef\expect#1{\left<\mathcode`\|"8000\let|\bravert {#1}\right>}}
\def\bravert{\egroup\,\vrule\,\bgroup}

\def\est{E_F^*}

\begin{document}

\title{The coherence-incoherence crossover and the
  mass-renormalization puzzles in \Sr2RuO4}

\author{Jernej~Mravlje}
\affiliation{Centre de Physique Th\'eorique, \'Ecole Polytechnique, CNRS,
91128 Palaiseau Cedex, France}
\affiliation{Jo\v{z}ef Stefan Institute, Jamova~39, Ljubljana, Slovenia}
\author{Markus~Aichhorn}
\affiliation{Institute of Theoretical and Computational Physics, TU Graz, Petersgasse 16, Graz, Austria}
\affiliation{Centre de Physique Th\'eorique, \'Ecole Polytechnique, CNRS,
91128 Palaiseau Cedex, France}
\author{Takashi~Miyake}
\affiliation{Nanosystem Research Institute, AIST, Tsukuba 305-8568, Japan}
\affiliation{Japan Science and Technology Agency, CREST, Kawaguchi
  332-0012, Japan}
\author{Kristjan~Haule}
\affiliation{Physics Department and Center for Materials Theory, Rutgers University, Piscataway NJ 08854, USA}
\author{Gabriel~Kotliar}
\affiliation{Physics Department and Center for Materials Theory, Rutgers University, Piscataway NJ 08854, USA}
\author{Antoine~Georges}
\affiliation{Centre de Physique Th\'eorique, \'Ecole Polytechnique, CNRS,
91128 Palaiseau Cedex, France}
\affiliation{Coll\`ege de France, 11 place Marcelin Berthelot, 75005 Paris, France}
\affiliation{Japan Science and Technology Agency, CREST, Kawaguchi
  332-0012, Japan}

\begin{abstract}
We calculate the electronic structure of \Sr2RuO4, treating
correlations in the framework of dynamical mean-field theory. The
approach successfully reproduces several experimental results and
explains the key properties of this material: the anisotropic mass
renormalization of quasiparticles and the crossover into an incoherent
regime at a low temperature. While the orbital differentiation
originates from the proximity of the van Hove singularity, strong
correlations are caused by the Hund's coupling. 
The generality of this mechanism for other correlated materials is pointed out. 
\end{abstract}

\maketitle
Fermi-liquid theory describes the low-energy excitations of metals
in terms of quasiparticles, which carry the
quantum numbers of a bare electron but have a renormalized mass $m^*$.
Quasiparticles have infinite lifetime on the Fermi surface and at temperature
$T=0$, but otherwise acquire a finite lifetime $\hbar/\Gamma$.
They carry only a fraction $Z$
of the total spectral weight associated with all single-particle excitations,
as encoded in the spectral function $A({\bf k},\omega)$.
A hallmark of strong correlations is that some
of these interaction-induced renormalizations
($m^*,Z^{-1},\Gamma$) become large.

The concept of a quasiparticle is meaningful only as long
as its inverse lifetime is smaller than the typical excitation or
thermal energy $\hbar\Gamma\lesssim \hbar\omega,kT$.
The internal consistency of Fermi-liquid theory rests on
$\hbar\Gamma \sim (kT)^2/\est \sim (\hbar\omega)^2/\est$,
due to phase-space constraints.
For temperatures larger than a coherence scale $T^*$ ($\sim\est/k$),
quasiparticles become short-lived and the Landau Fermi-liquid
description no longer applies.
Due to strong correlations,
$T^*$ can be much lower than the bare electronic scale $E_F/k$.
The description of the incoherent regime $T>T^*$ and of the associated crossover
is a major challenge which requires new concepts and techniques.
Of all transition metal oxides, the layered perovskite \Sr2RuO4 is
undoubtedly the one in which the Fermi liquid regime has been most
studied~\cite{mackenzie03,*bergemann03}.
Resistivities %
obey accurately a $T^2$ law for
$T\lesssim 30$K~\cite{hussey98}, despite the large anisotropy
$\rho_c/\rho_{ab}\sim 10^3$.  \Sr2RuO4 is also an ideal material to
investigate the crossover into the incoherent regime.
Indeed, at $130$K,
$\rho_c(T)$ reaches a maximum and decreases as temperature is further
increased, while the $T$-dependence of $\rho_{ab}$ remains metallic.
ARPES studies indicate that %
quasiparticle peaks %
disappear (by broadening and
loosing spectral weight) at a temperature
close to that where $\rho_c$ reaches its maximum~\cite{wang04,kidd05,*[{see also }] valla02}.

The 3-sheet Fermi surface of this material has been accurately
determined by quantum oscillation experiments~\cite{bergemann03} and
is reasonably well described by electronic structure calculations in the
local density approximation (LDA)~\cite{oguchi95,*singh95}.
On the other hand, the measured masses are not reproduced by the
LDA. Three bands of mainly $t_{2g}$ character cross the Fermi
surface. The broadest ($3.5$~eV) band of $xy$ character gives rise to a
two-dimensional Fermi surface sheet $\gamma$. The degenerate $xz$ and $yz$
orbitals give rise to narrower ($1.5$~eV) bands with quasi
one-dimensional Fermi surface sheets $\alpha$ and $\beta$.
Experimentally, large and {\it anisotropic} mass enhancements
$m^*/m_{\mathrm{LDA}}$ are found, namely $(3, 3.5, 5.5)$ for sheets
$\alpha, \beta, \gamma$, respectively~\cite{bergemann03}.

These experimental findings raise several puzzles, unresolved
to this day.
The large effective masses and the low %
coherence
scale indicate that \Sr2RuO4 is a strongly correlated material.
Surprisingly~\cite{konik07}, 
 the largest mass enhancement is actually observed for the widest ($xy$) band.
Furthermore, Ru being a $4d$ element, the screened on-site repulsion
is not expected to be large ($U\lesssim 3$~eV, somewhat smaller
than the bandwidth).
In a nutshell, these puzzles can be loosely summarized by the question:
why is \Sr2RuO4 strongly correlated ?

In this letter, we answer these questions in terms of the electronic
structure of the material.
Treating correlation effects within dynamical
mean-field theory (DMFT), we achieve
quantitative agreement with experiments. 
At a qualitative level, our explanation relies on the
Hund's coupling $J$ and the proximity of the van Hove singularity
for the $xy$ band.  These key elements of our picture,
especially the Hund's coupling, have general relevance
to $4d$ transition-metal oxides, as well as to other materials in
which strong correlation effects are observed but are not due to a strong
Hubbard $U$ or the proximity to a Mott insulator.

The calculations use the full potential implementation of LDA+DMFT as
presented in Ref.~\cite{aichhorn09}. The framework of Ref.~\cite{haule10}
gives very similar results. Wannier-like $t_{2g}$ orbitals are
constructed out of Kohn-Sham bands within the energy window $[-3,1]\,$eV
with respect to the Fermi energy.
We use the full rotationally invariant interaction appropriate for
a correct description of atomic multiplets:
\begin{eqnarray}
  H_{I} &=& U \sum_m n_{m\uparrow}  n_{m\downarrow} +
\sum_{m<n,\sigma}[U' n_{m\sigma} n_{n\bar{\sigma}}  \nonumber \\
      &+& (U'-J)n_{m\sigma}n_{n \sigma}  - J c_{m\sigma}^{\dagger}
    c_{m\bar{\sigma}} c_{n\bar{\sigma}}^{\dagger} c_{n\sigma}] \nonumber\\
    &-& J \sum_{m<n} [ c_{m\uparrow}^{\dagger}
    c^{\dagger}_{m\downarrow} c_{n\uparrow} c_{n\downarrow} + h.c. ]
\label{eq:hami}
\end{eqnarray}
where $J$ is the Hund's coupling constant, $U'=U-2J$ and $m,n$ run
over $t_{2g}$ orbitals. Ru $e_{g}$ and O $p$
orbitals are not explicitly included.
The importance of correlations leading to charge transfer among the orbitals,
mass renormalizations and satellites was
recognized in earlier studies~\cite{liebsch00,*anisimov02, *pchelkina07, *[{see also }]gorelov10}.
We use the strong-coupling continuous-time Monte-Carlo impurity
solver~\cite{werner06, *[{see also }]haule07} in order to reach the
low-temperature regime where the coherence-incoherence crossover takes
place~\footnote{
We use $>10^7$ Monte Carlo steps per iteration and $10^6$ steps for the thermalization. Worst average sign $>0.997$. }.
We calculated the interaction parameter $U$ from first-principles
using constrained-RPA~\cite{aryasetiawan04, *[{For technical details see }]miyake08}. The interaction matrix is
found to be quite isotropic with $U=2.5\,$eV for $xy$ and $U=2.2\,$eV
for $xz$ orbitals. The stronger mass enhancement of the $xy$ orbital can thus
not be explained by an anisotropy of the interactions~\cite{konik07}.

\begin{table}
\begin{ruledtabular}
\begin{tabular}{c c c c c c}
  $J$ [eV] & $m^*/m_{\rm{LDA}}|_{xy}$  & $m^*/m_{\rm{LDA}}|_{xz}$ &  $T_{xy}^*$[K] &  $T^*_{xz}$[K] & $T_>$[K]\\
\hline
  0.0, 0.1 & 1.7  & 1.7 & $>1000$ & $>1000$ & $>1000$ \\ %
  0.2 & 2.3  & 2.0 & 300 &  800 & $>1000$\\ %
  0.3 & 3.2  & 2.4 & 100 & 300 & 500\\  %
  0.4 & 4.5  & 3.3 & 60 & 150 & 350 \\  %
\end{tabular}
\end{ruledtabular}
\caption{\label{table1}
Mass enhancement of the $xy$ and $xz$ orbitals,
as a function of Hund's coupling, for $U=2.3$~eV.
Other columns: coherence temperatures as defined in the text.
}
\end{table}

\begin{figure}
 \begin{center}
\includegraphics[width=75mm,keepaspectratio]{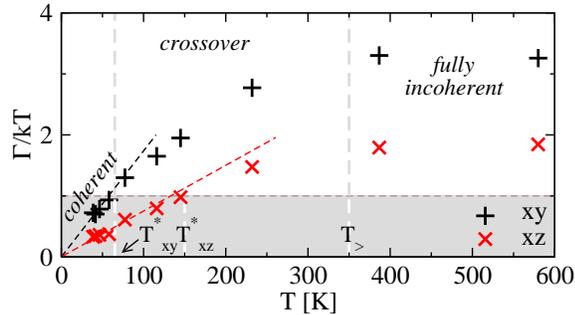}
   \end{center}
   \caption{\label{fig:z_sigmas}
    Temperature-dependence of $\Gamma/kT$, with $\hbar/\Gamma$ %
  the quasiparticle lifetime. %
   The shading indicates the `coherent' regime with long-lived
   quasiparticles such that $\Gamma\lesssim kT$.}
 \end{figure}

We now turn to results. In table~\ref{table1} we report the mass
enhancements of each orbital, given within DMFT by:
$m^*/m_\mathrm{LDA}=Z^{-1}\big|_{\,T\to 0}$ with $Z^{-1}=1-\partial\mathrm{Im}\Sigma(i\omega)/
\partial\omega\big|_{\omega \to 0^+}$.
 The derivative is
extracted by fitting a fourth-order polynomial to the data for
the lowest six Matsubara frequencies. 
The calculated mass enhancements for $U=2.3$~eV, $J=0.4$~eV (used in the remainder of the
paper~\footnote{
Extracting $J$ from the reduced $J_{mm'}$ matrix calculated by constrained RPA
yielded $J=0.25$eV, although a somewhat larger value is expected to
be obtained when considering the full $U_{m_1m_2m_3m_4}$ matrix.})
are found to be close to the experiment%
~\cite{mackenzie03,bergemann03}.

Table~\ref{table1} demonstrates that the Hund's coupling is essential
to reproduce the observed magnitude of mass enhancements and the $xy-xz$ differentiation.
A comparable mass enhancement (but without $xy-xz$ differentiation)
occurs at $J=0$ only for the unphysically large $U=5$~eV.
In addition we find that, by favoring maximal
angular momentum, the Hund's coupling drives the populations of orbitals closer to one another 
(to $1.29$ and $1.36$, for xy and xz, respectively) in comparison to the LDA value ($1.23,1.39$), hence
improving the agreement with quantum oscillations experiments ($\sim
1.33, 1.33$).

\begin{figure*}[tbp!]
 \begin{center}
\includegraphics[width=15.0cm]{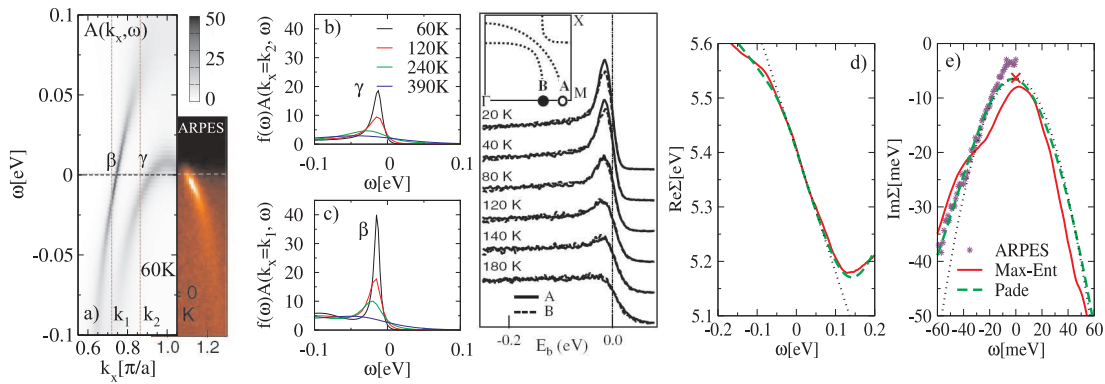}

   \end{center}
\caption{\label{fig:arpes_spectra}
(a) Intensity map of the spectral function $A(k,\omega)$ along $\Gamma\to$M
  for $0.55\pi/a\leq k_x\leq 1.05\pi/a$, $k_y=0$ at $T=60$~K compared to ARPES~\cite{shen07}. (b,c) Spectral lineshapes at wavevectors $k_1$, $k_2$ compared to ARPES~\cite{wang04}. (d,e)
  $\rm{Im}\Sigma(\omega+i0^+)$ and $\rm{Re}\Sigma(\omega+i0^+)$
   for xz orbital at $T=60$~K obtained by stochastic maximum entropy (plain line) and
   Pade approximants (dashed) compared to ARPES~\cite{ingle05}. Also indicated (cross,dotted line)
   are the low-$\omega$ behavior from a polynomial fit.}
\end{figure*}

To understand the coherence-incoherence crossover, we look at the
inverse quasiparticle lifetime, presented in Fig.~\ref{fig:z_sigmas}
as $\Gamma/kT$ {\it vs.} $T$, with $\Gamma=-Z\rm{Im}\Sigma(i0^+)$.  At
very low temperatures the Fermi-liquid $\Gamma\propto T^2$ behavior is
indicated (dashed).  We define the coherence scale $T^*$ by
$\Gamma(T^*)/kT^*=1$, but the deviations from $T^2$-law are visible
already at lower temperatures. $T^*$ is reported in Table~\ref{table1}
and also indicated on Fig.~\ref{fig:z_sigmas}. We see that $T^*$ is as
low as $60$~K for the most correlated $xy$ orbital.  At high
temperatures $T\gtrsim T_{>}\sim 400$~K, $\Gamma/kT$ saturates,
signaling the `incoherent' regime characterized by a quasi-linear
temperature dependence $\Gamma \propto kT$. An intermediate crossover
region where $\Gamma/kT$ gradually increases connects these two
regimes.

How do these regimes reveal themselves when probed by
spectroscopic experiments?  The left-most panel of
Fig.~\ref{fig:arpes_spectra} displays an intensity map of the
momentum-resolved spectral function demonstrating that
our results compare well with ARPES~\cite{shen07}.
Panels (b) and (c) display the
energy-distribution curves at two specific momenta. In the `coherent'
regime, these spectra display sharp peaks corresponding to the
Fermi surface crossings. Upon increasing
temperature the quasiparticle peaks broaden and above $T_>$
cannot be discerned anymore.
Note that in ARPES~\cite{wang04} the peaks disappear already at a somewhat
lower temperature, possibly due to the finite momentum resolution in
experiment.

The crossover scale $kT_>$ manifests itself also in the dependence 
of the self-energy on frequency, displayed in the rightmost panels of
Fig.~\ref{fig:arpes_spectra}.
We observe that deviations from the low-frequency Fermi liquid
regime $\rm{Re}\Sigma \sim \Sigma(0)+\omega(1-1/Z)$,
$\rm{Im}\Sigma\sim \omega^2 +(\pi T)^2$ appear at an
energy scale of order $40$~meV$\sim kT_>$, at which a  `kink'~\cite{byczuk07} is observed in $\rm{Re}\Sigma(\omega)$.
Such a feature at that energy scale is indeed reported in ARPES
 (Fig.~\ref{fig:arpes_spectra})~\cite{ingle05, aiura04}.

\begin{figure}[h]
 \begin{center}
\includegraphics[width=75mm,keepaspectratio]{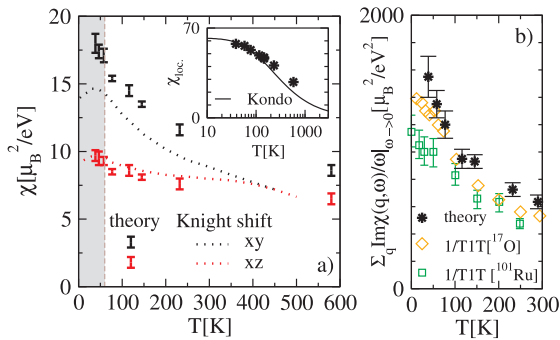}
   \end{center}
   \caption{\label{fig:nmr} (a) The uniform susceptibility $\chi_m (q=0)$ for each orbital,
     $m=xy,xz$.  (Inset) Total local susceptibility
     $\chi_{\mathrm{loc.}}=\sum_{qm}\chi_m(q)$ compared to that of the
     S=1 Kondo model. (b) $\lim_{\omega\to0}\sum_q{\mathrm{Im} \chi(q,
       w)}/\omega$ compared  
     to %
     NMR~\cite{imai98}. }
 \end{figure}

The crossover also affects the
magnetic response.  On Fig.~\ref{fig:nmr}(a) we display the
orbitally resolved uniform magnetic susceptibilities and compare them to
the NMR Knight shift measurements~\cite{imai98}.
Saturation to a Pauli magnetic susceptibility is observed only
below $T^*$ (shaded). The stronger temperature dependence of the xy
orbital related to $T^*_{xy} < T^*_{xz}$ is reproduced well.
The total low temperature uniform susceptibility (1.2 emu/mol) is
within the estimated error ($\sim30\%$~\cite{imai98}) of the
thermodynamic measurements (0.9 emu/mol)~\cite{maeno97})~\footnote{The 
NMR estimates (dots) are thus multiplied by $1.3$, to ease the
comparison of the temperature dependence}.
We also calculated the local susceptibility (inset) and found that
it is larger than the uniform one, especially for the xz orbital
\footnote{$\chi_m(q=0)^{-1}-[\sum_q \chi_m(q)]^{-1}$, averaged over the
  temperature interval, are $0.06,0.2$ $\mathrm{eV}/(g \mu_b)^2$, for
  xy, xz, resp}. This signals {\it antiferromagnetic}
  correlations, in agreement with experimental
  observations~\cite{sidis99}.  

In Fig.~\ref{fig:nmr}(b) we display
$\lim_{\omega\to0}\sum_q{\mathrm{Im} \chi(q,\omega)}/\omega$ and
compare to the NMR data for $1/T_1T$ (we used the 
values of hyperfine couplings from Refs.~\cite{sidis99, ishida01b}).  There, the data
saturate only well below $T^*$, illustrating that the Fermi liquid
behavior in two-particle properties is more fragile than in
single-particle ones. Indeed, in the well known Kondo problem the
Kondo resonance persists at temperatures up to $2 T_K$ while the
magnetic susceptibility saturates to a Pauli form only below $T_K/5$.

Having demonstrated that the LDA+DMFT results agree with
experimental data, we turn to theoretical insights.
In DMFT the local physics is revealed by solving an impurity
model (atom + bath) with atomic 
interactions given by Eq.~(\ref{eq:hami}).
This can be rewritten as $H_I=(U-3J) n (n-1)/2 - 2J S^2
-(J/2) T^2$, where $S$ is the total spin and $T$ is the total angular
momentum~\cite{leo04thesis}.
 The four-electrons subspace
separates into five $T=2,S=0$ states, a single $T=0,S=0$ state and nine $T=1,S=1$
states. At $J=0$ all these states are degenerate and constitute a
$15$-dimensional representation of an $SU(6)$ symmetry group. This high
degeneracy results in a very high  coherence
scale $\sim 0.5$eV and small mass renormalizations (see Table~\ref{table1}).
The Hund's coupling $J$ lowers the SU(6) symmetry down to
$\mathrm{SU}(2)_\mathrm{spin} \times \mathrm{SU}(2)_{\mathrm{orbit}}$
with the $9$-fold degenerate atomic multiplet $S=1,T=1$ having lowest energy.
The ground state of the impurity model is non-degenerate with
$S=0,T=0$ corresponding to exact screening of this atomic
multiplet~\cite{leo04thesis}. Thus \Sr2RuO4 is a Fermi liquid.
The Hund's coupling projects the spin degrees of freedom
to a low energy manifold characterized by a reduced Kondo coupling,
resulting in a suppressed Kondo scale~\cite{okada73, nevidomskiy09}.
The effective low energy model is in our case a $S=1$ Kondo
model. Indeed, the inset of Fig.~\ref{fig:nmr}(a) demonstrates that
at low $T$ the LDA+DMFT result for $\chi_{\mathrm{loc.}}$ is fit well 
by the $S=1$ Kondo model Bethe ansatz curve \cite{[{A single fit parameter $T_K=240$K in the notation of }] desgranges85}.

The dramatic reduction of coherence scale as a result of the Hund's
coupling has been noted before in impurity
models~\cite{okada73,pruschke05,nevidomskiy09}, DMFT studies of model
Hamiltonians~\cite{werner09} and for iron
pnictides~\cite{haule09,*[{see also }]aichhorn10}.  It occurs whenever
multiplet correlations persist while the on-site $U$ is strongly
screened (due to the large spatial extension of the correlated orbital
as in $4d$ transition metal oxides, or the large polarizability of
screening orbitals as in pnictides).

\begin{figure}[h]
 \begin{center}
\includegraphics[width=55mm,keepaspectratio]{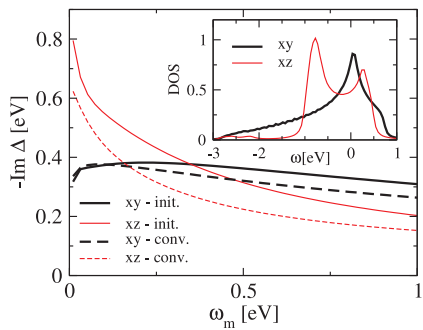}
   \end{center}
   \caption{\label{fig:dos_hyb} The hybridization functions $\Delta(i
     \omega)$ at the initial DMFT step and at self
     consistency. (Inset) The LDA projected density of states. 
}
 \end{figure}

The origin of the larger $xy$ effective mass can be traced to the
proximity of the van-Hove singularity.
Higher density of states near the Fermi level implies weaker dispersion
and in turn reflects in a lower value of the respective hybridization function $\Delta(i \omega)$ at
low frequencies (Fig.~\ref{fig:dos_hyb}).  Indeed, ignoring the
self-consistency (i.e. on the first DMFT iteration),
$\mathrm{Im}{\Delta}^{(1)}(i0^+)=-\pi\rho_F/\left[\mathrm{Re}{G}_{\mathrm{loc}}(i0^{+})^2+
  (\pi\rho_F)^2\right]\simeq -1/(\pi\rho_F)$ with $\rho_F$ the LDA
density of states at the Fermi level.  The large value of $\rho_F$
thus corresponds to a suppressed low-energy effective
hopping~\cite{[{A single band calculation with the orbitally
  projected density of states at filling $4/3$ gives $m^*/m\sim
  2.2(1.8)$ and $T^*=0.05(0.1)$eV for xy (xz). See also }] zitko09, *schmitt10}.  In contrast, the
full bandwidth is larger for the $xy$, and so is the LDA kinetic
energy ($0.27$eV for $xy$, $0.20$eV for $xz$).  This reflects in the 
high-frequency behavior of the hybridization, indeed larger for $xy$
at high-frequency.  Note that the degree of correlation cannot
be guessed from the kinetic energy or bandwidth of each band, which
would naively suggest a smaller mass for $xy$, in contrast to
observations.

In summary, we have demonstrated that several experimental results for
\Sr2RuO4 are well reproduced by the LDA+DMFT method.  We have shown
that the suppression of the coherence scale is due to the Hund's
coupling, and pointed out the generality of this mechanism.  We have
also shown that the orbital differentiation and larger $xy$ mass is
due to the difference in low-energy hybridization properties of each
orbital, caused by their orientation-dependent bonding properties in
this anisotropic material.  This is expected to be relevant to other
layered perovskites, most notably to the metal-insulator transition in
\CaSrRuO4. 

\acknowledgements{We are grateful to P.~Bourges, M.~Fabrizio,
  M.~Ferrero, E.~Gull, L.~de~Leo, A.~Mackenzie, Y.~Sidis and
  M. Sigrist for useful discussions.  We acknowledge the support of
  the NSF-materials world network (NSF DMR 0806937), the Partner
  University Fund, the CNRS-LIA program, the Austrian Science Fund
  (projects J2760, F4103, P18551) and the hospitality of KITP, Santa
  Barbara (NSF PHY05-51164) and of CPHT (G.K.).}

\bibliography{sr2ruo4}

\begin{thebibliography}{41}%
\makeatletter
\providecommand \@ifxundefined [1]{%
 \@ifx{#1\undefined}
}%
\providecommand \@ifnum [1]{%
 \ifnum #1\expandafter \@firstoftwo
 \else \expandafter \@secondoftwo
 \fi
}%
\providecommand \@ifx [1]{%
 \ifx #1\expandafter \@firstoftwo
 \else \expandafter \@secondoftwo
 \fi
}%
\providecommand \natexlab [1]{#1}%
\providecommand \enquote  [1]{``#1''}%
\providecommand \bibnamefont  [1]{#1}%
\providecommand \bibfnamefont [1]{#1}%
\providecommand \citenamefont [1]{#1}%
\providecommand \href@noop [0]{\@secondoftwo}%
\providecommand \href [0]{\begingroup \@sanitize@url \@href}%
\providecommand \@href[1]{\@@startlink{#1}\@@href}%
\providecommand \@@href[1]{\endgroup#1\@@endlink}%
\providecommand \@sanitize@url [0]{\catcode `\\12\catcode `\$12\catcode
  `\&12\catcode `\#12\catcode `\^12\catcode `\_12\catcode `\%12\relax}%
\providecommand \@@startlink[1]{}%
\providecommand \@@endlink[0]{}%
\providecommand \url  [0]{\begingroup\@sanitize@url \@url }%
\providecommand \@url [1]{\endgroup\@href {#1}{\urlprefix }}%
\providecommand \urlprefix  [0]{URL }%
\providecommand \Eprint [0]{\href }%
\providecommand \doibase [0]{http://dx.doi.org/}%
\providecommand \selectlanguage [0]{\@gobble}%
\providecommand \bibinfo  [0]{\@secondoftwo}%
\providecommand \bibfield  [0]{\@secondoftwo}%
\providecommand \translation [1]{[#1]}%
\providecommand \BibitemOpen [0]{}%
\providecommand \bibitemStop [0]{}%
\providecommand \bibitemNoStop [0]{.\EOS\space}%
\providecommand \EOS [0]{\spacefactor3000\relax}%
\providecommand \BibitemShut  [1]{\csname bibitem#1\endcsname}%
\let\auto@bib@innerbib\@empty
\bibitem [{\citenamefont {Mackenzie}\ and\ \citenamefont
  {Maeno}(2003)}]{mackenzie03}%
  \BibitemOpen
  \bibfield  {author} {\bibinfo {author} {\bibfnamefont {A.~P.}\ \bibnamefont
  {Mackenzie}}\ and\ \bibinfo {author} {\bibfnamefont {Y.}~\bibnamefont
  {Maeno}},\ }\href {\doibase 10.1103/RevModPhys.75.657} {\bibfield  {journal}
  {\bibinfo  {journal} {Rev. Mod. Phys.}\ }\textbf {\bibinfo {volume} {75}},\
  \bibinfo {pages} {657} (\bibinfo {year} {2003})}\BibitemShut {NoStop}%
\bibitem [{\citenamefont {Bergemann}\ \emph {et~al.}(2003)\citenamefont
  {Bergemann} \emph {et~al.}}]{bergemann03}%
  \BibitemOpen
  \bibfield  {author} {\bibinfo {author} {\bibfnamefont {C.}~\bibnamefont
  {Bergemann}} \emph {et~al.},\ }\href@noop {} {\bibfield  {journal} {\bibinfo
  {journal} {Adv. Phys.}\ }\textbf {\bibinfo {volume} {52}},\ \bibinfo {pages}
  {639} (\bibinfo {year} {2003})}\BibitemShut {NoStop}%
\bibitem [{\citenamefont {Hussey}\ \emph {et~al.}(1998)\citenamefont {Hussey}
  \emph {et~al.}}]{hussey98}%
  \BibitemOpen
  \bibfield  {author} {\bibinfo {author} {\bibfnamefont {N.~E.}\ \bibnamefont
  {Hussey}} \emph {et~al.},\ }\href {\doibase 10.1103/PhysRevB.57.5505}
  {\bibfield  {journal} {\bibinfo  {journal} {Phys. Rev. B}\ }\textbf {\bibinfo
  {volume} {57}},\ \bibinfo {pages} {5505} (\bibinfo {year}
  {1998})}\BibitemShut {NoStop}%
\bibitem [{\citenamefont {Wang}\ \emph {et~al.}(2004)\citenamefont {Wang} \emph
  {et~al.}}]{wang04}%
  \BibitemOpen
  \bibfield  {author} {\bibinfo {author} {\bibfnamefont {S.-C.}\ \bibnamefont
  {Wang}} \emph {et~al.},\ }\href {\doibase 10.1103/PhysRevLett.92.137002}
  {\bibfield  {journal} {\bibinfo  {journal} {Phys. Rev. Lett.}\ }\textbf
  {\bibinfo {volume} {92}},\ \bibinfo {pages} {137002} (\bibinfo {year}
  {2004})}\BibitemShut {NoStop}%
\bibitem [{\citenamefont {Kidd}\ \emph {et~al.}(2005)\citenamefont {Kidd} \emph
  {et~al.}}]{kidd05}%
  \BibitemOpen
  \bibfield  {author} {\bibinfo {author} {\bibfnamefont {T.~E.}\ \bibnamefont
  {Kidd}} \emph {et~al.},\ }\href {\doibase 10.1103/PhysRevLett.94.107003}
  {\bibfield  {journal} {\bibinfo  {journal} {Phys. Rev. Lett.}\ }\textbf
  {\bibinfo {volume} {94}},\ \bibinfo {pages} {107003} (\bibinfo {year}
  {2005})}\BibitemShut {NoStop}%
\bibitem [{\citenamefont {Valla}\ \emph {et~al.}(2002)\citenamefont {Valla}
  \emph {et~al.}}]{valla02}%
  \BibitemOpen
  \bibfield  {author} {\bibinfo {author} {\bibfnamefont {T.}~\bibnamefont
  {Valla}} \emph {et~al.},\ }\href@noop {} {\bibfield  {journal} {\bibinfo
  {journal} {Nature}\ }\textbf {\bibinfo {volume} {417}},\ \bibinfo {pages}
  {627} (\bibinfo {year} {2002})}\BibitemShut {NoStop}%
\bibitem [{\citenamefont {Oguchi}(1995)}]{oguchi95}%
  \BibitemOpen
  \bibfield  {author} {\bibinfo {author} {\bibfnamefont {T.}~\bibnamefont
  {Oguchi}},\ }\href {\doibase 10.1103/PhysRevB.51.1385} {\bibfield  {journal}
  {\bibinfo  {journal} {Phys. Rev. B}\ }\textbf {\bibinfo {volume} {51}},\
  \bibinfo {pages} {1385} (\bibinfo {year} {1995})}\BibitemShut {NoStop}%
\bibitem [{\citenamefont {Singh}(1995)}]{singh95}%
  \BibitemOpen
  \bibfield  {author} {\bibinfo {author} {\bibfnamefont {D.~J.}\ \bibnamefont
  {Singh}},\ }\href {\doibase 10.1103/PhysRevB.52.1358} {\bibfield  {journal}
  {\bibinfo  {journal} {Phys. Rev. B}\ }\textbf {\bibinfo {volume} {52}},\
  \bibinfo {pages} {1358} (\bibinfo {year} {1995})}\BibitemShut {NoStop}%
\bibitem [{\citenamefont {Konik}\ and\ \citenamefont {Rice}(2007)}]{konik07}%
  \BibitemOpen
  \bibfield  {author} {\bibinfo {author} {\bibfnamefont {R.~M.}\ \bibnamefont
  {Konik}}\ and\ \bibinfo {author} {\bibfnamefont {T.~M.}\ \bibnamefont
  {Rice}},\ }\href {\doibase 10.1103/PhysRevB.76.104501} {\bibfield  {journal}
  {\bibinfo  {journal} {Phys. Rev. B}\ }\textbf {\bibinfo {volume} {76}},\
  \bibinfo {pages} {104501} (\bibinfo {year} {2007})}\BibitemShut {NoStop}%
\bibitem [{\citenamefont {Aichhorn}\ \emph {et~al.}(2009)\citenamefont
  {Aichhorn} \emph {et~al.}}]{aichhorn09}%
  \BibitemOpen
  \bibfield  {author} {\bibinfo {author} {\bibfnamefont {M.}~\bibnamefont
  {Aichhorn}} \emph {et~al.},\ }\href {\doibase 10.1103/PhysRevB.80.085101}
  {\bibfield  {journal} {\bibinfo  {journal} {Phys. Rev. B}\ }\textbf {\bibinfo
  {volume} {80}},\ \bibinfo {pages} {085101} (\bibinfo {year}
  {2009})}\BibitemShut {NoStop}%
\bibitem [{\citenamefont {Haule}\ \emph {et~al.}(2010)\citenamefont {Haule}
  \emph {et~al.}}]{haule10}%
  \BibitemOpen
  \bibfield  {author} {\bibinfo {author} {\bibfnamefont {K.}~\bibnamefont
  {Haule}} \emph {et~al.},\ }\href {\doibase 10.1103/PhysRevB.81.195107}
  {\bibfield  {journal} {\bibinfo  {journal} {Phys. Rev. B}\ }\textbf {\bibinfo
  {volume} {81}},\ \bibinfo {pages} {195107} (\bibinfo {year}
  {2010})}\BibitemShut {NoStop}%
\bibitem [{\citenamefont {Liebsch}\ and\ \citenamefont
  {Lichtenstein}(2000)}]{liebsch00}%
  \BibitemOpen
  \bibfield  {author} {\bibinfo {author} {\bibfnamefont {A.}~\bibnamefont
  {Liebsch}}\ and\ \bibinfo {author} {\bibfnamefont {A.}~\bibnamefont
  {Lichtenstein}},\ }\href@noop {} {\bibfield  {journal} {\bibinfo  {journal}
  {Phys. Rev. Lett.}\ }\textbf {\bibinfo {volume} {84}},\ \bibinfo {pages}
  {1591} (\bibinfo {year} {2000})}\BibitemShut {NoStop}%
\bibitem [{\citenamefont {Anisimov}\ \emph {et~al.}(2002)\citenamefont
  {Anisimov} \emph {et~al.}}]{anisimov02}%
  \BibitemOpen
  \bibfield  {author} {\bibinfo {author} {\bibfnamefont {V.}~\bibnamefont
  {Anisimov}} \emph {et~al.},\ }\href@noop {} {\bibfield  {journal} {\bibinfo
  {journal} {Eur. Phys. J. B}\ }\textbf {\bibinfo {volume} {25}},\ \bibinfo
  {pages} {191} (\bibinfo {year} {2002})}\BibitemShut {NoStop}%
\bibitem [{\citenamefont {Pchelkina}\ \emph {et~al.}(2007)\citenamefont
  {Pchelkina} \emph {et~al.}}]{pchelkina07}%
  \BibitemOpen
  \bibfield  {author} {\bibinfo {author} {\bibfnamefont {Z.~V.}\ \bibnamefont
  {Pchelkina}} \emph {et~al.},\ }\href@noop {} {\bibfield  {journal} {\bibinfo
  {journal} {Phys. Rev. B}\ }\textbf {\bibinfo {volume} {75}},\ \bibinfo
  {pages} {035122} (\bibinfo {year} {2007})}\BibitemShut {NoStop}%
\bibitem [{\citenamefont {Gorelov}\ \emph {et~al.}(2010)\citenamefont
  {Gorelov}, \citenamefont {Karolak}, \citenamefont {Wehling}, \citenamefont
  {Lechermann}, \citenamefont {Lichtenstein},\ and\ \citenamefont
  {Pavarini}}]{gorelov10}%
  \BibitemOpen
  \bibfield  {author} {\bibinfo {author} {\bibfnamefont {E.}~\bibnamefont
  {Gorelov}}, \bibinfo {author} {\bibfnamefont {M.}~\bibnamefont {Karolak}},
  \bibinfo {author} {\bibfnamefont {T.~O.}\ \bibnamefont {Wehling}}, \bibinfo
  {author} {\bibfnamefont {F.}~\bibnamefont {Lechermann}}, \bibinfo {author}
  {\bibfnamefont {A.~I.}\ \bibnamefont {Lichtenstein}}, \ and\ \bibinfo
  {author} {\bibfnamefont {E.}~\bibnamefont {Pavarini}},\ }\href {\doibase
  10.1103/PhysRevLett.104.226401} {\bibfield  {journal} {\bibinfo  {journal}
  {Phys. Rev. Lett.}\ }\textbf {\bibinfo {volume} {104}},\ \bibinfo {pages}
  {226401} (\bibinfo {year} {2010})}\BibitemShut {NoStop}%
\bibitem [{\citenamefont {Werner}\ \emph {et~al.}(2006)\citenamefont {Werner}
  \emph {et~al.}}]{werner06}%
  \BibitemOpen
  \bibfield  {author} {\bibinfo {author} {\bibfnamefont {P.}~\bibnamefont
  {Werner}} \emph {et~al.},\ }\href {\doibase 10.1103/PhysRevLett.97.076405}
  {\bibfield  {journal} {\bibinfo  {journal} {Phys. Rev. Lett.}\ }\textbf
  {\bibinfo {volume} {97}},\ \bibinfo {pages} {076405} (\bibinfo {year}
  {2006})}\BibitemShut {NoStop}%
\bibitem [{\citenamefont {Haule}(2007)}]{haule07}%
  \BibitemOpen
  \bibfield  {author} {\bibinfo {author} {\bibfnamefont {K.}~\bibnamefont
  {Haule}},\ }\href {\doibase 10.1103/PhysRevB.75.155113} {\bibfield  {journal}
  {\bibinfo  {journal} {Phys. Rev. B}\ }\textbf {\bibinfo {volume} {75}},\
  \bibinfo {pages} {155113} (\bibinfo {year} {2007})}\BibitemShut {NoStop}%
\bibitem [{Note1()}]{Note1}%
  \BibitemOpen
  \bibinfo {note} {We use $>10^7$ Monte Carlo steps per iteration and $10^6$
  steps for the thermalization. Worst average sign $>0.997$.}\BibitemShut
  {Stop}%
\bibitem [{\citenamefont {Aryasetiawan}\ \emph {et~al.}(2004)\citenamefont
  {Aryasetiawan} \emph {et~al.}}]{aryasetiawan04}%
  \BibitemOpen
  \bibfield  {author} {\bibinfo {author} {\bibfnamefont {F.}~\bibnamefont
  {Aryasetiawan}} \emph {et~al.},\ }\href {\doibase 10.1103/PhysRevB.70.195104}
  {\bibfield  {journal} {\bibinfo  {journal} {Phys. Rev. B}\ }\textbf {\bibinfo
  {volume} {70}},\ \bibinfo {pages} {195104} (\bibinfo {year}
  {2004})}\BibitemShut {NoStop}%
\bibitem [{\citenamefont {Miyake}\ and\ \citenamefont
  {Aryasetiawan}(2008)}]{miyake08}%
  \BibitemOpen
  \bibfield  {author} {\bibinfo {author} {\bibfnamefont {T.}~\bibnamefont
  {Miyake}}\ and\ \bibinfo {author} {\bibfnamefont {F.}~\bibnamefont
  {Aryasetiawan}},\ }\href {\doibase 10.1103/PhysRevB.77.085122} {\bibfield
  {journal} {\bibinfo  {journal} {Phys. Rev. B}\ }\textbf {\bibinfo {volume}
  {77}},\ \bibinfo {pages} {085122} (\bibinfo {year} {2008})}\BibitemShut
  {NoStop}%
\bibitem [{Note2()}]{Note2}%
  \BibitemOpen
  \bibinfo {note} {Extracting $J$ from the reduced $J_{mm'}$ matrix calculated
  by constrained RPA yielded $J=0.25$eV, although a somewhat larger value is
  expected to be obtained when considering the full $U_{m_1m_2m_3m_4}$
  matrix.}\BibitemShut {Stop}%
\bibitem [{\citenamefont {Shen}\ \emph {et~al.}(2007)\citenamefont {Shen} \emph
  {et~al.}}]{shen07}%
  \BibitemOpen
  \bibfield  {author} {\bibinfo {author} {\bibfnamefont {K.~M.}\ \bibnamefont
  {Shen}} \emph {et~al.},\ }\href {\doibase 10.1103/PhysRevLett.99.187001}
  {\bibfield  {journal} {\bibinfo  {journal} {Phys. Rev. Lett.}\ }\textbf
  {\bibinfo {volume} {99}},\ \bibinfo {pages} {187001} (\bibinfo {year}
  {2007})}\BibitemShut {NoStop}%
\bibitem [{\citenamefont {Ingle}\ \emph {et~al.}(2005)\citenamefont {Ingle}
  \emph {et~al.}}]{ingle05}%
  \BibitemOpen
  \bibfield  {author} {\bibinfo {author} {\bibfnamefont {N.~J.~C.}\
  \bibnamefont {Ingle}} \emph {et~al.},\ }\href {\doibase
  10.1103/PhysRevB.72.205114} {\bibfield  {journal} {\bibinfo  {journal} {Phys.
  Rev. B}\ }\textbf {\bibinfo {volume} {72}},\ \bibinfo {pages} {205114}
  (\bibinfo {year} {2005})}\BibitemShut {NoStop}%
\bibitem [{\citenamefont {Byczuk}\ \emph {et~al.}(2007)\citenamefont {Byczuk}
  \emph {et~al.}}]{byczuk07}%
  \BibitemOpen
  \bibfield  {author} {\bibinfo {author} {\bibfnamefont {K.}~\bibnamefont
  {Byczuk}} \emph {et~al.},\ }\href@noop {} {\bibfield  {journal} {\bibinfo
  {journal} {Nat. Phys.}\ }\textbf {\bibinfo {volume} {3}},\ \bibinfo {pages}
  {168} (\bibinfo {year} {2007})}\BibitemShut {NoStop}%
\bibitem [{\citenamefont {Aiura}\ \emph {et~al.}(2004)\citenamefont {Aiura}
  \emph {et~al.}}]{aiura04}%
  \BibitemOpen
  \bibfield  {author} {\bibinfo {author} {\bibfnamefont {Y.}~\bibnamefont
  {Aiura}} \emph {et~al.},\ }\href {\doibase 10.1103/PhysRevLett.93.117005}
  {\bibfield  {journal} {\bibinfo  {journal} {Phys. Rev. Lett.}\ }\textbf
  {\bibinfo {volume} {93}},\ \bibinfo {pages} {117005} (\bibinfo {year}
  {2004})}\BibitemShut {NoStop}%
\bibitem [{\citenamefont {Imai}\ \emph {et~al.}(1998)\citenamefont {Imai} \emph
  {et~al.}}]{imai98}%
  \BibitemOpen
  \bibfield  {author} {\bibinfo {author} {\bibfnamefont {T.}~\bibnamefont
  {Imai}} \emph {et~al.},\ }\href {\doibase 10.1103/PhysRevLett.81.3006}
  {\bibfield  {journal} {\bibinfo  {journal} {Phys. Rev. Lett.}\ }\textbf
  {\bibinfo {volume} {81}},\ \bibinfo {pages} {3006} (\bibinfo {year}
  {1998})}\BibitemShut {NoStop}%
\bibitem [{\citenamefont {Maeno}\ \emph {et~al.}(1997)\citenamefont {Maeno}
  \emph {et~al.}}]{maeno97}%
  \BibitemOpen
  \bibfield  {author} {\bibinfo {author} {\bibfnamefont {Y.}~\bibnamefont
  {Maeno}} \emph {et~al.},\ }\href@noop {} {\bibfield  {journal} {\bibinfo
  {journal} {J. Phys. Soc. Jpn}\ }\textbf {\bibinfo {volume} {66}},\ \bibinfo
  {pages} {1405} (\bibinfo {year} {1997})}\BibitemShut {NoStop}%
\bibitem [{Note3()}]{Note3}%
  \BibitemOpen
  \bibinfo {note} {The NMR estimates (dots) are thus multiplied by $1.3$, to
  ease the comparison of the temperature dependence}\BibitemShut {NoStop}%
\bibitem [{Note4()}]{Note4}%
  \BibitemOpen
  \bibinfo {note} {$\chi _m(q=0)^{-1}-[\DOTSB \sum@ \slimits@ _q \chi
  _m(q)]^{-1}$, averaged over the temperature interval, are $0.06,0.2$
  $\protect \mathrm {eV}/(g \mu _b)^2$, for xy, xz, resp}\BibitemShut {NoStop}%
\bibitem [{\citenamefont {Sidis}\ \emph {et~al.}(1999)\citenamefont {Sidis}
  \emph {et~al.}}]{sidis99}%
  \BibitemOpen
  \bibfield  {author} {\bibinfo {author} {\bibfnamefont {Y.}~\bibnamefont
  {Sidis}} \emph {et~al.},\ }\href {\doibase 10.1103/PhysRevLett.83.3320}
  {\bibfield  {journal} {\bibinfo  {journal} {Phys. Rev. Lett.}\ }\textbf
  {\bibinfo {volume} {83}},\ \bibinfo {pages} {3320} (\bibinfo {year}
  {1999})}\BibitemShut {NoStop}%
\bibitem [{\citenamefont {Ishida}\ \emph {et~al.}(2001)\citenamefont {Ishida}
  \emph {et~al.}}]{ishida01b}%
  \BibitemOpen
  \bibfield  {author} {\bibinfo {author} {\bibfnamefont {K.}~\bibnamefont
  {Ishida}} \emph {et~al.},\ }\href@noop {} {\bibfield  {journal} {\bibinfo
  {journal} {Phys. Rev. B}\ }\textbf {\bibinfo {volume} {64}},\ \bibinfo
  {pages} {100501} (\bibinfo {year} {2001})}\BibitemShut {NoStop}%
\bibitem [{\citenamefont {de~Leo}(2004)}]{leo04thesis}%
  \BibitemOpen
  \bibfield  {author} {\bibinfo {author} {\bibfnamefont {L.}~\bibnamefont
  {de~Leo}},\ }\href@noop {} {Ph.D. thesis},\ \bibinfo  {school} {SISSA}
  (\bibinfo {year} {2004})\BibitemShut {NoStop}%
\bibitem [{\citenamefont {Okada}\ and\ \citenamefont {Yosida}(1973)}]{okada73}%
  \BibitemOpen
  \bibfield  {author} {\bibinfo {author} {\bibfnamefont {I.}~\bibnamefont
  {Okada}}\ and\ \bibinfo {author} {\bibfnamefont {K.}~\bibnamefont {Yosida}},\
  }\href@noop {} {\bibfield  {journal} {\bibinfo  {journal} {Prog. Theor.
  Phys.}\ }\textbf {\bibinfo {volume} {49}},\ \bibinfo {pages} {1483} (\bibinfo
  {year} {1973})}\BibitemShut {NoStop}%
\bibitem [{\citenamefont {Nevidomskyy}\ and\ \citenamefont
  {Coleman}(2009)}]{nevidomskiy09}%
  \BibitemOpen
  \bibfield  {author} {\bibinfo {author} {\bibfnamefont {A.~H.}\ \bibnamefont
  {Nevidomskyy}}\ and\ \bibinfo {author} {\bibfnamefont {P.}~\bibnamefont
  {Coleman}},\ }\href {\doibase 10.1103/PhysRevLett.103.147205} {\bibfield
  {journal} {\bibinfo  {journal} {Phys. Rev. Lett.}\ }\textbf {\bibinfo
  {volume} {103}},\ \bibinfo {pages} {147205} (\bibinfo {year}
  {2009})}\BibitemShut {NoStop}%
\bibitem [{\citenamefont {Desgranges}(1985)}]{desgranges85}%
  \BibitemOpen
  \bibfield  {author} {\bibinfo {author} {\bibfnamefont {H.-U.}\ \bibnamefont
  {Desgranges}},\ }\href@noop {} {\bibfield  {journal} {\bibinfo  {journal} {J.
  Phys. C}\ }\textbf {\bibinfo {volume} {18}},\ \bibinfo {pages} {5481}
  (\bibinfo {year} {1985})}\BibitemShut {NoStop}%
\bibitem [{\citenamefont {Pruschke}\ and\ \citenamefont
  {Bulla}(2005)}]{pruschke05}%
  \BibitemOpen
  \bibfield  {author} {\bibinfo {author} {\bibfnamefont {T.}~\bibnamefont
  {Pruschke}}\ and\ \bibinfo {author} {\bibfnamefont {R.}~\bibnamefont
  {Bulla}},\ }\href@noop {} {\bibfield  {journal} {\bibinfo  {journal} {Eur.
  Phys. J. B}\ }\textbf {\bibinfo {volume} {44}},\ \bibinfo {pages} {217}
  (\bibinfo {year} {2005})}\BibitemShut {NoStop}%
\bibitem [{\citenamefont {Werner}\ \emph {et~al.}(2009)\citenamefont {Werner},
  \citenamefont {Gull},\ and\ \citenamefont {Millis}}]{werner09}%
  \BibitemOpen
  \bibfield  {author} {\bibinfo {author} {\bibfnamefont {P.}~\bibnamefont
  {Werner}}, \bibinfo {author} {\bibfnamefont {E.}~\bibnamefont {Gull}}, \ and\
  \bibinfo {author} {\bibfnamefont {A.~J.}\ \bibnamefont {Millis}},\ }\href
  {\doibase 10.1103/PhysRevB.79.115119} {\bibfield  {journal} {\bibinfo
  {journal} {Phys. Rev. B}\ }\textbf {\bibinfo {volume} {79}},\ \bibinfo
  {pages} {115119} (\bibinfo {year} {2009})}\BibitemShut {NoStop}%
\bibitem [{\citenamefont {Haule}\ and\ \citenamefont
  {Kotliar}(2009)}]{haule09}%
  \BibitemOpen
  \bibfield  {author} {\bibinfo {author} {\bibfnamefont {K.}~\bibnamefont
  {Haule}}\ and\ \bibinfo {author} {\bibfnamefont {G.}~\bibnamefont
  {Kotliar}},\ }\href@noop {} {\bibfield  {journal} {\bibinfo  {journal} {New
  J. Phys.}\ }\textbf {\bibinfo {volume} {11}},\ \bibinfo {pages} {025021}
  (\bibinfo {year} {2009})}\BibitemShut {NoStop}%
\bibitem [{\citenamefont {Aichhorn}\ \emph {et~al.}(2010)\citenamefont
  {Aichhorn} \emph {et~al.}}]{aichhorn10}%
  \BibitemOpen
  \bibfield  {author} {\bibinfo {author} {\bibfnamefont {M.}~\bibnamefont
  {Aichhorn}} \emph {et~al.},\ }\href@noop {} {\bibfield  {journal} {\bibinfo
  {journal} {Phys. Rev. B}\ }\textbf {\bibinfo {volume} {82}},\ \bibinfo
  {pages} {064504} (\bibinfo {year} {2010})}\BibitemShut {NoStop}%
\bibitem [{\citenamefont {\ifmmode~\check{Z}\else \v{Z}\fi{}itko}\ \emph
  {et~al.}(2009)\citenamefont {\ifmmode~\check{Z}\else \v{Z}\fi{}itko},
  \citenamefont {Bon\ifmmode~\check{c}\else \v{c}\fi{}a},\ and\ \citenamefont
  {Pruschke}}]{zitko09}%
  \BibitemOpen
  \bibfield  {author} {\bibinfo {author} {\bibfnamefont {R.}~\bibnamefont
  {\ifmmode~\check{Z}\else \v{Z}\fi{}itko}}, \bibinfo {author} {\bibfnamefont
  {J.}~\bibnamefont {Bon\ifmmode~\check{c}\else \v{c}\fi{}a}}, \ and\ \bibinfo
  {author} {\bibfnamefont {T.}~\bibnamefont {Pruschke}},\ }\href {\doibase
  10.1103/PhysRevB.80.245112} {\bibfield  {journal} {\bibinfo  {journal} {Phys.
  Rev. B}\ }\textbf {\bibinfo {volume} {80}},\ \bibinfo {pages} {245112}
  (\bibinfo {year} {2009})}\BibitemShut {NoStop}%
\bibitem [{\citenamefont {Schmitt}(2010)}]{schmitt10}%
  \BibitemOpen
  \bibfield  {author} {\bibinfo {author} {\bibfnamefont {S.}~\bibnamefont
  {Schmitt}},\ }\href {\doibase 10.1103/PhysRevB.82.155126} {\bibfield
  {journal} {\bibinfo  {journal} {Phys. Rev. B}\ }\textbf {\bibinfo {volume}
  {82}},\ \bibinfo {pages} {155126} (\bibinfo {year} {2010})}\BibitemShut
  {NoStop}%
\end{thebibliography}%

\end{document}